\documentclass[11pt]{article}

\usepackage{arxiv}

\usepackage[utf8]{inputenc} 
\usepackage[T1]{fontenc}    
\usepackage{hyperref}       
\usepackage{url}            
\usepackage{booktabs}       
\usepackage{tabularx}
\usepackage{amsfonts}       
\usepackage{nicefrac}       
\usepackage{microtype}      
\usepackage{lipsum}		
\usepackage{graphicx}
\usepackage[
  backend=biber,
  style=numeric-comp,
  sorting=none,
  sortcites=true,
  maxnames=5, minnames=3
]{biblatex}
\usepackage{doi}

\usepackage{xeCJK}
\usepackage{enumitem}
\usepackage{mdframed}
\usepackage{minted}
\setlist[itemize,enumerate]{leftmargin=1.5em}

\title{UCAgent: An End-to-End Agent for Block-Level Functional Verification}

\date{} 					

\author{
  \textbf{Junyue Wang \textsuperscript{1,2}} \qquad
  \textbf{Zhicheng Yao \textsuperscript{1,2}} \qquad
  \textbf{Yan Pi \textsuperscript{1,2}}\qquad
  \textbf{Xiaolong Li\textsuperscript{1,2}} \qquad
  \textbf{Fangyuan Song\textsuperscript{3}} \\
  \textbf{Jinru Wang\textsuperscript{2}} \qquad
  \textbf{Yunlong Xie\textsuperscript{1,2}} \qquad
  \textbf{Sa Wang\textsuperscript{\dag,1,2}} \qquad
  \textbf{Yungang Bao\textsuperscript{1,2,3}} \\
  \textsuperscript{1} State Key Lab of Processors, Institute of Computing Technology, CAS \\
  \textsuperscript{2} University of Chinese Academy of Sciences \\
  \textsuperscript{3} Beijing Institute of Open Source Chip \\
  \texttt{wangjunyue24@mails.ucas.ac.cn}
}

\renewcommand{\headeright}{}
\renewcommand{\undertitle}{}

\hypersetup{
pdftitle={UCAgent: An End-to-End Agent for Block-Level Functional Verification},
pdfsubject={cs.AI, cs.AR},
pdfauthor={Junyue Wang},
pdfkeywords={Chip Verification, Functional Verification,Large Language Models, Agent, Automated Verification, Verification Consistency},
}

\begin{document}
\maketitle

\begin{abstract}
Functional verification remains a critical bottleneck in modern IC development cycles, accounting for approximately 70\% of total development time in many projects.
However, traditional methods, including constrained-random and formal verification, struggle to keep pace with the growing complexity of modern semiconductor designs.

While recent advances in Large Language Models (LLMs) have shown promise in code generation and task automation, significant challenges hinder the realization of end-to-end functional verification automation.
These challenges include (i) limited accuracy in generating Verilog/SystemVerilog verification code, (ii) the fragility of LLMs when executing complex, multi-step verification workflows, and (iii) the difficulty of maintaining verification consistency across specifications, coverage models, and test cases throughout the workflow.

To address these challenges, we propose UCAgent, an end-to-end agent that automates hardware block-level functional verification based on three core mechanisms.
First, we establish a pure Python verification environment using Picker and Toffee to avoid relying on LLM-generated SystemVerilog verification code.
Second, we introduce a configurable 31-stage fine-grained verification workflow to guide the LLM, where each stage is verified by an automated checker.
Furthermore, we propose a Verification Consistency Labeling Mechanism (VCLM) that assigns hierarchical labels to LLM-generated artifacts, improving the reliability and traceability of verification.

Experimental results show that UCAgent can complete end-to-end automated verification on multiple modules, including the UART, FPU, and integer divider modules, achieving up to 98.5\% code coverage and up to 100\% functional coverage. UCAgent also discovers previously unidentified design defects in realistic designs, demonstrating its practical potential.
\end{abstract}


\section{Introduction}

Functional verification is the critical process of ensuring that a hardware design's implementation correctly realizes its intended specification, making it the most resource-intensive phase of the IC design\cite{molina2007functional}.

As hardware designs grow exponentially in complexity, however, verification has become a server bottleneck in terms of time, labor, and reliability.
Industry studies indicate that verification now consumes over 70\% of the total IC project timeline~\cite{daniel2023state}. 
Consequently, the demand for verification engineers significantly outpaces that for design engineers, while designers themselves must spend nearly half of their work time to assist with verification~\cite{foster2022fcstudy}. 
Yet, these massive investments fail to guarantee reliable outcomes. 
In 2024, only 14\% of projects achieved first-silicon success, and logic/functional defects remain the leading cause of costly re-spins~\cite{foster2024fcstudy}, underscoring the urgent need for a highly automated verification paradigm.

The industry has historically relied on constrained random verification~\cite{Mehta2018}, formal verification~\cite{kern1999formal}, and hardware emulation. However, these techniques have merely maintained a fragile balance rather than fundamentally closing the gap between design complexity and verification capability~\cite{foster2025bottleneck}.

Against this backdrop, both industry and academia have viewed AI-driven automated verification as a promising path to breaking the bottleneck.
Major EDA vendors (e.g., Synopsys~\cite{synopsys_ai,synopsys_vso_ai}, Cadence~\cite{cadence_jedai, cadence_verisium, cadence_verisium_debug}, and Siemens~\cite{siemens_eda_ai_system, siemens_solido_ai}) have integrated AI into various stages of the verification flow, including assertion generation, coverage closure, and debugging. Concurrently, academia has explored the protential of using LLMs for specific tasks such as assertion generation~\cite{rahul2024llmassertions, sun2023nl2sva, liu2024domainadapted, pulavarthi2025ready}, automated testbench generation~\cite{bhandari2024fsmtestbench, huang2024verilogassistant, xiao2024neuroverify}, and coverage improvement~\cite{bhandari2024fsmtestbench, huang2024verilogassistant, xiao2024neuroverify}.
Yet, these approaches mainly optimize \textit{specific stages} of the verification process mentioned above, rather than automating the end-to-end verification process. Engineers are still required to orchestrate and integrate the distinct stages.

Recent work has demonstrated the potential of LLMs for automated functional verification. Based on the verification language, these efforts can be broadly categorized into two paradigms:
\begin{itemize}[topsep=0pt, partopsep=0pt]
    \item \textbf{SystemVerilog \& UVM-based approaches:} UVM²~\cite{ye2025uvm2} uses a hybrid generation strategy combining templates with LLMs to automatically generate UVM testbenches from design specifications and iteratively optimizes testcases using coverage feedback; MAVF~\cite{liu2025multi} proposes a multi-agent collaboration architecture that decouples verification into multiple specialized stages, including specification parsing, verification planning, testbench specification generation, and code generation, with specialized agents executing them in sequence and validating effectiveness on real chip modules.
    \item \textbf{Python-based approaches:} PRO-V~\cite{zhao2025pro} adopts a pure-Python generation strategy and achieves end-to-end RTL automated verification by combining it with an LLM-as-a-judge mechanism.
\end{itemize}

While these efforts confirm that LLM-based verification can significantly reduce time and labor costs, they still exhibit limitations in workflow flexibility, execution reliability, and result traceability.
To build an autonomous agent system that spans the entire verification flow, three core challenges must be addressed:

\textbf{Challenge 1: Limited accuracy of LLM-generated Verilog/SystemVerilog verification code.} LLMs are less proficient at generating hardware description languages (HDLs; e.g., Verilog/SystemVerilog) compared to software languages like Python. This stems primarily from two factors:
\begin{itemize}[topsep=0pt, partopsep=0pt]
    \item \textbf{Severe disparity in training data.} LLMs rely on massive datasets for pattern learning, and data scale directly affects model's performance. For example, in The Stack v2~\cite{lozhkov2024starcoder}, the largest open code dataset to date, Python accounts for 178.44 GB of data, while Verilog and SystemVerilog account for only 9.348 GB and 0.84 GB, respectively. This scarcity results in worse and less stable generation quality for HDLs.
    \item \textbf{Limited generation accuracy.} Consequently, LLMs exhibit limited accuracy in Verilog/SystemVerilog and struggle with complex syntactic correctness~\cite{tsai2024rtlfixer,  zhang2025understanding}. Although many studies~\cite{cui2024origen,zhao2025codev,ho2025verilogcoder,zhu2025codev} have worked to improve generation quality, most focus on generating hardware design code. As a result, even state-of-the-art models struggle to consistently and stably generate syntactically correct and semantically consistent SystemVerilog verification code~\cite{pinckney2025comprehensive}.
\end{itemize}

\textbf{Challenge 2: Difficulty completing complex end-to-end verification tasks.} 
The inherent complexity of functional verification requires processing massive volumes of design specifications, design/verification code and simulation reports, a sheer scale of information that easily exceeds the effective context window of current LLMs. 
In this setting, inherent issues of LLMs such as hallucination~\cite{enwiki:1316197198}, context rot~\cite{hong2025context}, and weak instruction following~\cite{heo2024llms} are amplified, severely impacting the reliability and effectiveness of the verification outcomes.
These limitations are particularly prominent in two aspects:
\begin{itemize}[topsep=0pt, partopsep=0pt]
    \item \textbf{Information loss and error accumulation in complex task execution.} Because functional verification stages are highly interdependent, any errors introduced in early phases (e.g., omitted constraints due to context loss, or syntactic flaws caused by hallucinations) accumulate rapidly and compromise the correctness of all subsequent stages. For instance, in existing work, MAVF requires multiple rounds of manual review and correction during testbench specification generation just to ensure later code generation accuracy. Therefore, an automated checking and feedback mechanism is strictly needed to constrain and correct outputs at each individual stage.
    \item \textbf{Inadequacy of simple workflows for diverse verification scenarios.} 
    To manage the inherent complexity of functional verification, tasks are typically decomposed into sequential stages for LLM execution. 
    However, the complexity of hardware modules varies significantly, rendering overly simple workflows ineffective across diverse scenarios. 
    For relatively simple designs (e.g., FIFOs or arithmetic units), interaction patterns are straightforward, allowing LLMs to directly generate stimuli and check output. 
    Conversely, complex modules with strict ordering and dependency constraints (e.g., bus crossbars) require the generation of pseudo-modules (mocks~\cite{osherove2024art}) to interact properly with the DUT. 
    Rudimentary workflows struggle with such sophisticated needs, often producing implementations with interface mismatches or behavioral deviations. 
    For instance, existing frameworks like UVM² rely on predefined generation sequences that, while stable for standard interfaces, lack the dynamic adaptability required to orchestrate complex hardware interactions. 
    Therefore, an effective and configurable workflow design is essential to ensure the successful completion of verification tasks across diverse scenarios
\end{itemize}

\textbf{Challenge 3: Maintaining verification consistency across the full flow is difficult.}
A robust functional verification process relies on strict traceability spanning the \textit{functional specification}, the \textit{coverage model}, and the \textit{testcase implementation}.
Specifically, every test point extracted from the specification must unambiguously map to corresponding covergroups and coverpoints, which are subsequently exercised by simulation stimuli.

We formally define this unbroken semantic correspondence across verification goals, implementations, and results as \textit{Verification Consistency}.
It is the foundation for building a verification closed loop and ensuring traceability of results.
However, LLMs often struggle to maintain this consistency.

\begin{figure}[h]
    \centering
    \includegraphics[width=\linewidth]{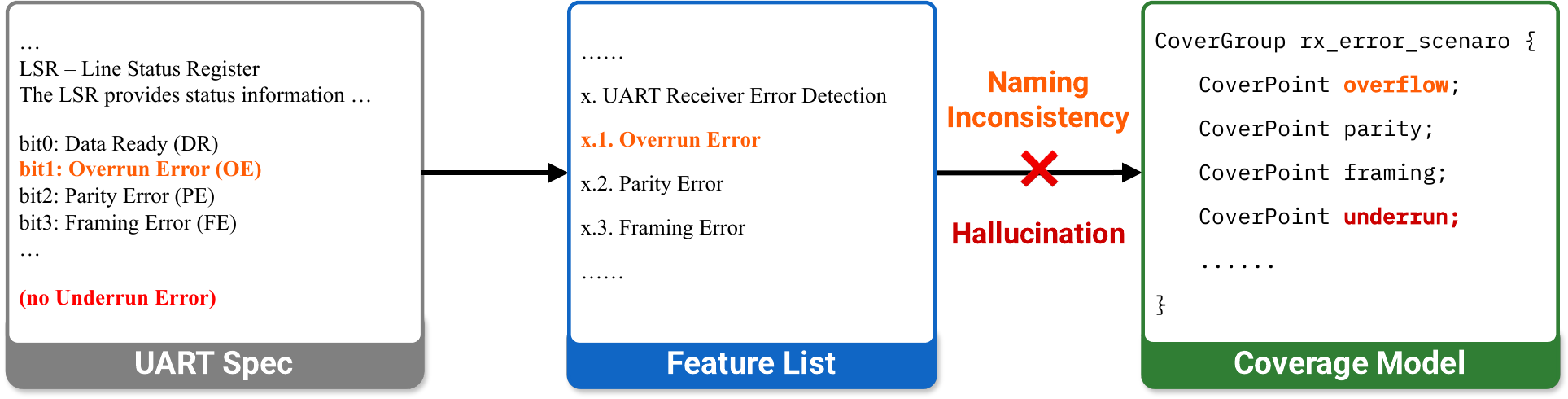}
    \caption{An example that violates verification consistency. The term \texttt{overrun} is replaced with the semantically similar \texttt{overflow}; additionally, an \texttt{underrun} scenario is incorrectly added. Such a scenario typically does not appear in UART designs and is more common in modules like SPI and USB host controllers~\cite{intel_xhci_reqspec_2019_rev122}.}
    \label{fig:vcmm-example}
\end{figure}

To illustrate this challenge, we use Qwen3-Coder-Plus to automate the translation from design specification into a feature list, and subsequently to a coverage model. 
As shown in Fig.~\ref{fig:vcmm-example}, two critical consistency failures emerge during this process. 
First, even when prompts explicitly require preserving the original terminology from the design specification, the LLM arbitrarily alters terms across these stages, causing severe naming inconsistencies. Second, the model hallucinates invalid verification scenarios that violate the actual design intent. Such failures sever the traceability, rendering the verification results unreliable.

Existing methods remain insufficient in maintaining verification consistency.
Specifically, PRO-V lacks a functional coverage model, failing to establish traceability from specification requirements to coverage metrics.
UVM² relies solely on prompt-based constraints, lacking explicit mechanisms to guarantee output compliance.
While MAVF builds a cross-reference matrix between test points and testcases, it lacks a comprehensive mechanism to sustain this consistency across the entire end-to-end verification flow.

To address these challenges, we propose UCAgent, an end-to-end agent that automates hardware block-level functional verification based on three core mechanisms.

First, we establish a pure-Python environment for the LLM to complete the verification tasks.
Specifically, we utilize Picker~\cite{picker} to convert the DUT into a Python package \texttt{PyDUT}, and leverage the Toffee~\cite{toffee} to guide the LLM in constructing testbenches and generating testcases entirely in Python.
By transforming the verification implementation from HDL development to Python generation, empowering the LLM to handle all verification activities efficiently.

Second, we design a 31-stage fine-grained verification workflow to guide the LLM.
It decomposes the verification process into manageable steps, spanning from feature extraction to coverage modeling, testcase generation, and bug analysis.
To mitigate inherent LLM issues such as context rot and hallucination, each stage incorporates automated checkers to validate results before the next stage.
This step-by-step feedback mechanism effectively controls error accumulation, ensuring reliable and quality-controlled execution.

Third, we propose a \textit{Verification Consistency Labeling Mechanism (VCLM)} to ensure unbroken traceability across the entire workflow. 
By requiring the LLM to assign hierarchical labels to generated artifacts, including specifications, coverage models, and testcases, VCLM formalizes implicit semantic correspondences into explicit, machine-checkable constraints. 
This formalization maintains strict alignment with the original design intent.

We evaluate UCAgent across diverse hardware modules, including UART, integer dividers, and ICache-WayLookup. 
Experimental results demonstrate that the agent successfully drives the entire verification lifecycle autonomously, achieving up to 98.5\% code coverage and 100\% functional coverage. 
Furthermore, UCAgent uncovers previously unidentified design defects in real-world verification scenarios, strongly validating the effectiveness and practical value of our proposed paradigm.

\section{Design Overview} 

This section introduces the three core mechanisms integrated into UCAgent to address the above challenges. These mechanisms enable fully automated end-to-end functional verification while ensuring reliability.

\begin{figure}[h]
    \centering
    \includegraphics[width=\textwidth]{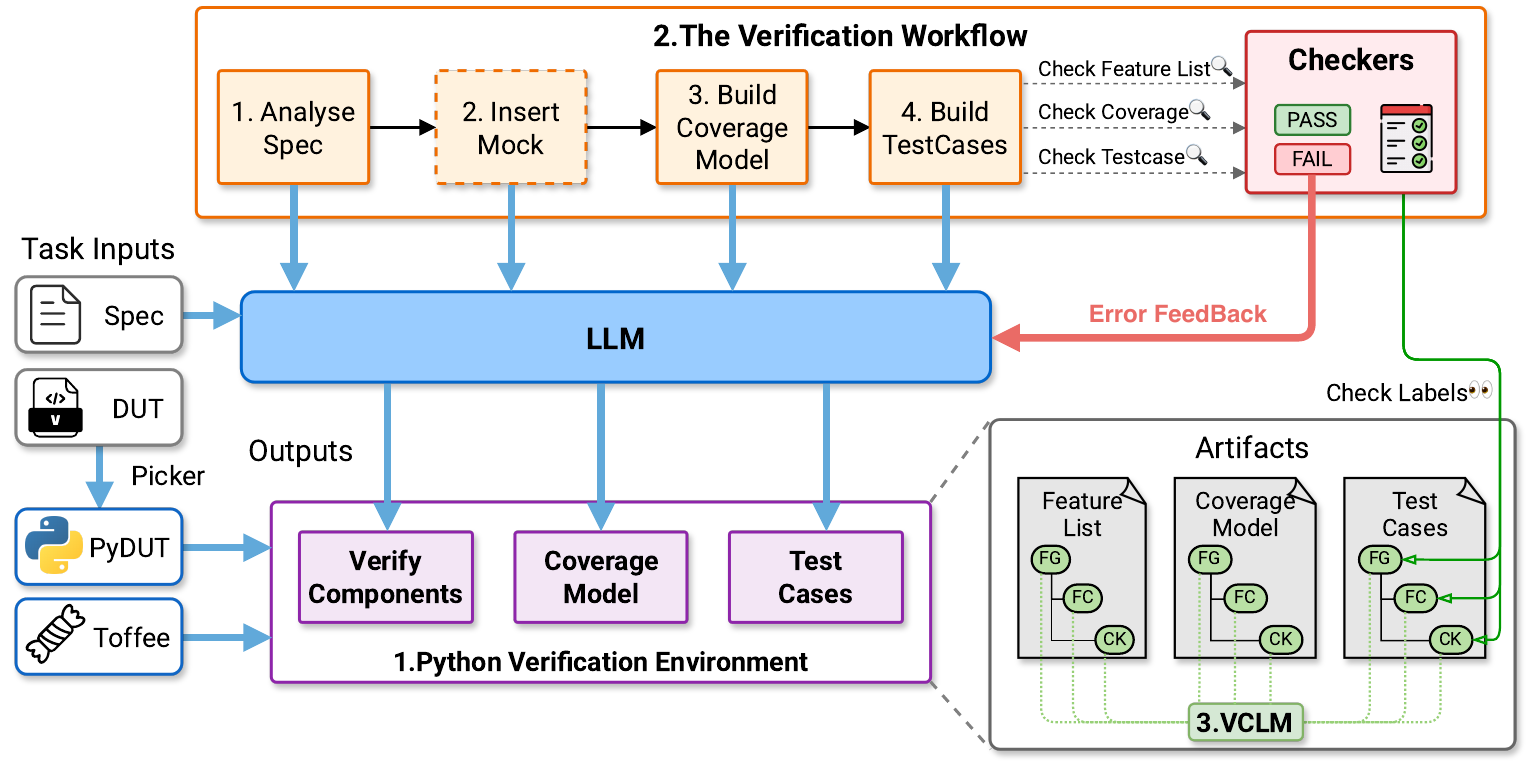}
    \caption{Relationships among UCAgent's core mechanisms}
    \label{fig:mechanism}
\end{figure}

\subsection{Python Verification Environment}

\noindent To bypass the data-scarcity bottleneck of hardware description languages, UCAgent establishes a pure-Python environment for the LLM to complete end-to-end verification tasks.
The primary limitation in current LLM-driven verification is the inability to reliably generate SystemVerilog (SV) verification code.
While SV is the industry standard, the scarcity of high-quality SV corpora in pre-training datasets prevents LLMs from mastering its complex semantics.
In contrast, LLMs demonstrate robust code-generation proficiency in Python~\cite{yang2025multiswebench}.

Within this environment, we utilize Picker~\cite{picker} to convert the DUT into a Python package (\textit{PyDUT}), and leverage the Toffee to provide high-level verification abstractions. Picker is a verification assistance tool that converts RTL designs described in Verilog/SystemVerilog into libraries or packages in high-level programming languages. Tofee is a Python-based verification framework that provides higher-level features and abstractions on the top of Picker.

This infrastructure abstracts hardware interactions, such as clock control, into simpler interfaces and significantly reduces dependency on simulation scheduling details like delta-cycles~\cite{cocotb_issue3110}.
By shielding the LLM from these low-level timing intricacies, the agent can reliably drive the hardware using lightweight in-context learning~\cite{brown2020language}.

The LLM interacts with \textit{PyDUT} via Toffee to construct testbenches and generate testcases.
This execution is orchestrated by our fine-grained workflow and validated by stage-specific checkers alongside the Verification Consistency Labeling Mechanism (VCLM).
Together, these components ensure syntactic correctness and maintain strict alignment with the design specifications, as illustrated in Fig.~\ref{fig:mechanism}.

\subsection{The Verification Workflow}
\begin{figure}[ht]
    \centering
    \includegraphics[width=\textwidth]{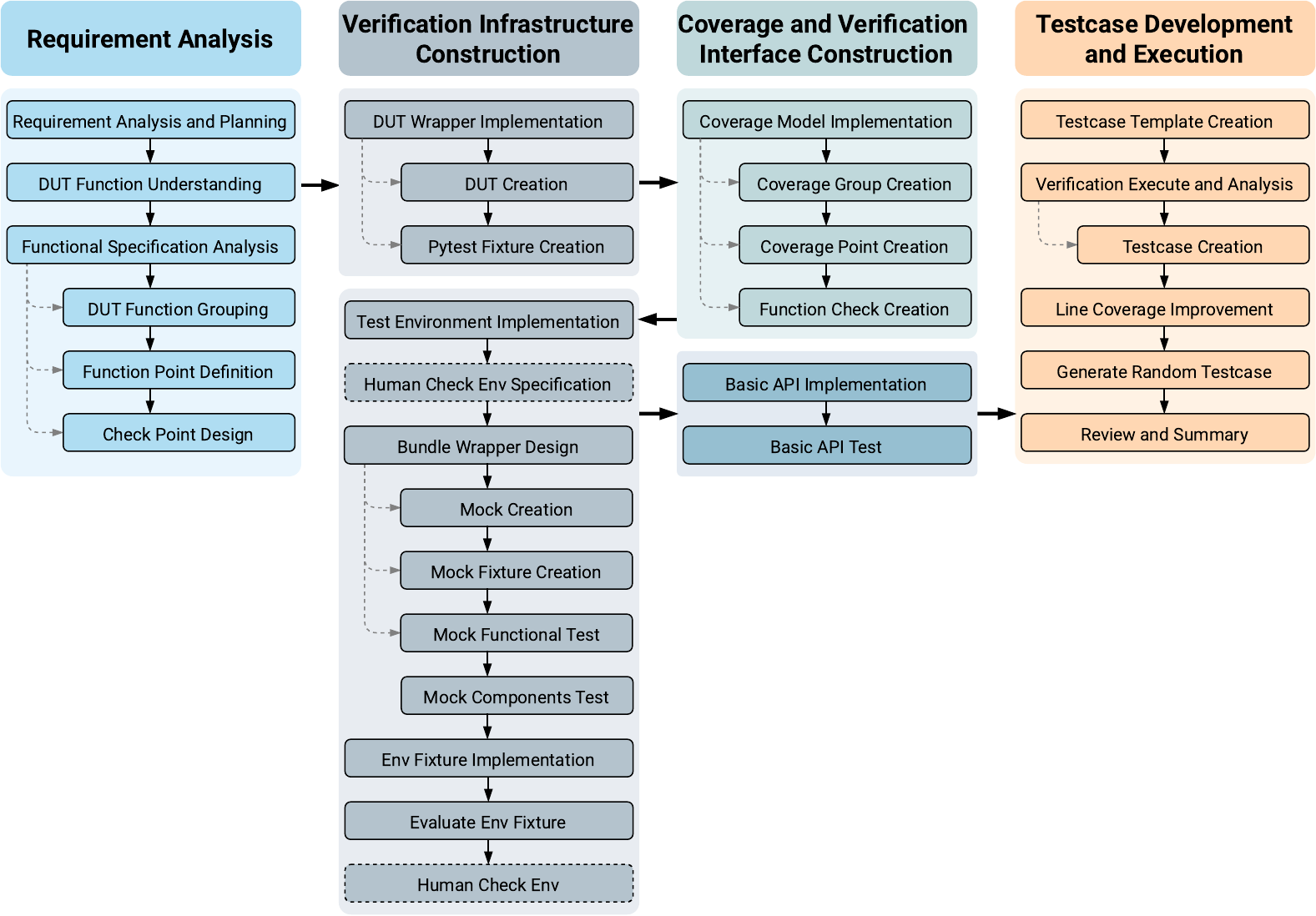}
    \caption{The detailed verification workflow. Dotted stages indicate that they are skipped by default}
    \label{fig:workflow}
\end{figure}

\noindent To manage the inherent complexity of hardware verification, UCAgent orchestrates the end-to-end process through a structured workflow.
It decomposes the verification lifecycle into fine-grained stages and enforces strict quality control by assigning a dedicated automated checker to each stage.

These stages cover the following aspects of the verification process:
\begin{itemize}[topsep=0pt, partopsep=0pt]
    \item \textbf{Requirement analysis and functional decomposition}, where the DUT behavior is interpreted and organized into structured verification targets.
    \item \textbf{Verification infrastructure construction}, where the DUT, fixtures, signal wrappers, and an optional mock-based environment are prepared.
    \item \textbf{Coverage and verification interface construction}, where functional intent is translated into executable coverage groups, checkpoints, and callable APIs.
    \item \textbf{Testcase development and execution}, where test templates are instantiated, refined into executable tests, and further extended with bug-oriented analysis and optional randomized testing.
\end{itemize}

The detailed stage breakdown is presented in Fig.~\ref{fig:workflow}.

The workflow splits the verification process into relatively independent stages, each with explicit inputs, deterministic outputs, and checkable deliverables.
For example, specification analysis must output a document with a precise label structure; coverage modeling must yield executable covergroup definitions; and testcase generation must produce simulation code that passes specific checks.

To validate these deliverables, the checker mechanism encodes expert quality criteria into executable constraint logic.
Upon the completion of a stage, a dedicated checker evaluates the generated artifact against predefined rules.
If the output fails to meet expectations, the checker returns concrete error messages to the LLM, guiding the model to iteratively self-correct and improving generation reliability.

This 31-stage architecture was derived empirically by refining the workflow based on checker feedback.
By analyzing checker-triggered retries and iteratively subdividing error-prone tasks, we ultimately converged on a fine-grained workflow that substantially reduced task complexity at each stage.

Finally, to accommodate the diverse complexities of different hardware modules, this workflow is designed to be fully configurable.
For instance, for modules with complex dependencies, a dedicated mock-generation stage can be inserted to replace the components on which the DUT depends.
This configurable execution model ensures both efficiency and reliability across varied verification scenarios.

\subsection{Verification Consistency Labeling Mechanism}
\noindent To address the difficulty of maintaining verification consistency across multiple stages, we propose the Verification Consistency Labeling Mechanism (VCLM). VCLM establishes a hierarchical labeling system comprising three distinct levels: \texttt{FG}, \texttt{FC}, and \texttt{CK}. UCAgent requires the LLM to consistently annotate these labels within the generated artifacts across three pivotal stages: functional specification analysis, coverage model construction, and testcase implementation.

By embedding these labels into the outputs, VCLM transforms implicit design dependencies into structured, machine-checkable data. Stage-specific checkers systematically extract these labels to validate cross-stage consistency. If a generated artifact lacks the required labels or contains labels that fail to match those established in preceding stages, the checker rejects the output. It then returns precise error traces to the LLM, triggering an immediate self-correction cycle to restore traceability.

Relying solely on natural language prompts, such as instructing the LLM to "ensure the coverage model is consistent with the specification," is inherently unreliable for complex hardware designs. In contrast, VCLM provides an explicit, formal carrier for verification intent. This structural constraint enforces strict behavioral boundaries on the LLM, fundamentally improving the reliability and traceability of the automated verification process, as illustrated in Fig.~\ref{fig:mechanism}.

\section{UCAgent Implementation}
\begin{figure}[tb]
    \centering
    \includegraphics[width=\textwidth]{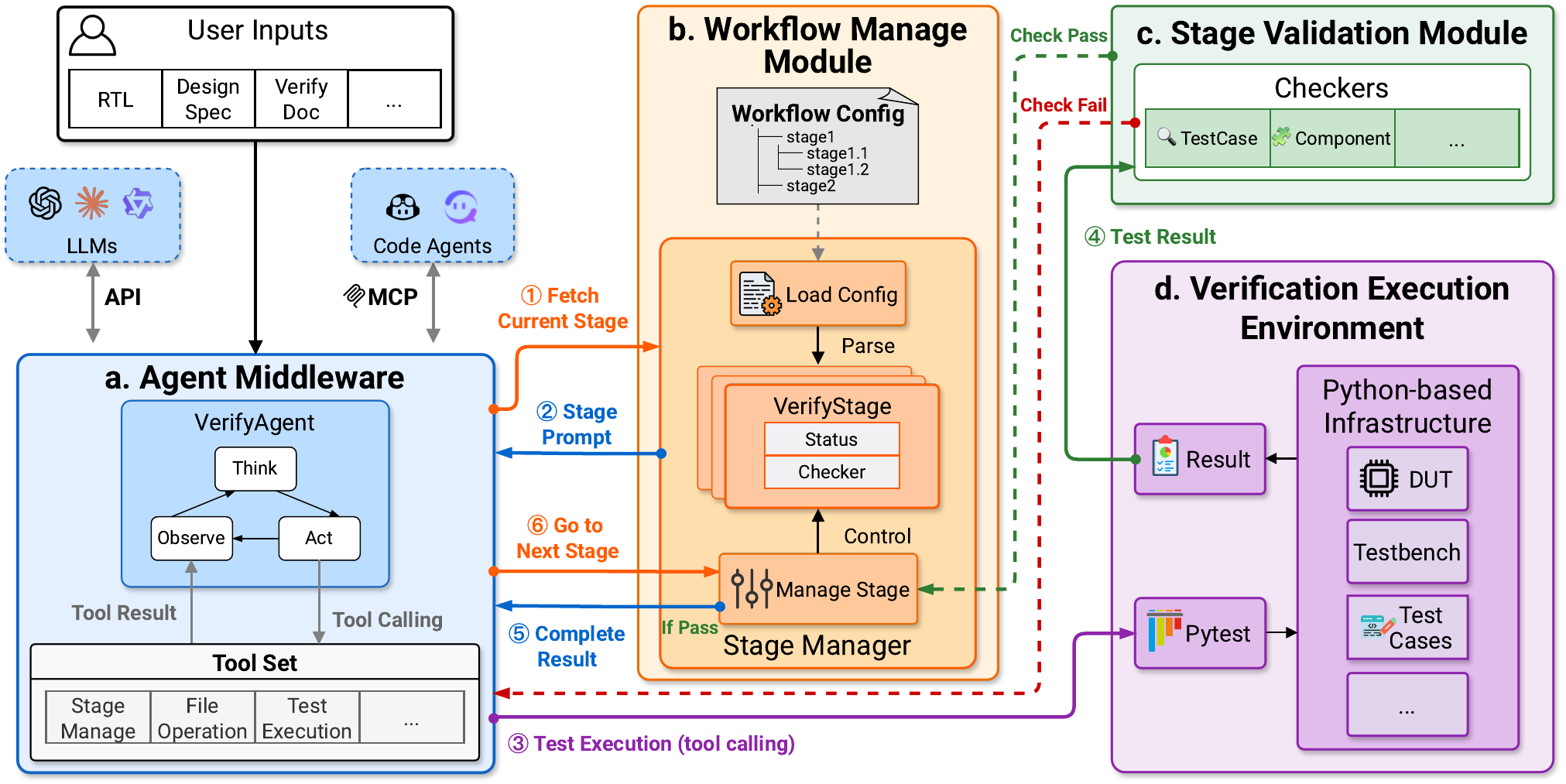}
    \caption{UCAgent implementation overview}
    \label{fig:overview}
\end{figure}

Based on the proposed mechanisms, we implement the UCAgent system.
As shown in Fig.~\ref{fig:overview}, UCAgent adheres to the principle of separation of concerns and comprises four independent modules.

The \textit{Agent Middleware} manages LLM interactions, context construction, and task reasoning.
The \textit{Workflow Management Module} orchestrates the execution sequence of the configurable 31-stage workflow.
The \textit{Stage Validation Module} integrates the automated checkers and VCLM to ensure output quality and cross-stage consistency.
Finally, the \textit{Verification Execution Environment} provides the Python-based infrastructure for test execution and hardware simulation.

In the remainder of this section, we detail these four core modules in turn.

\subsection{Agent Middleware}
\noindent The \textit{Agent Middleware} bridges the LLM and the verification environment.
Its primary function is to translate LLM reasoning into executable actions via tool calling~\cite{abdelaziz2024granitefunctioncallingmodelintroducing}.
Built upon the LangChain framework~\cite{langchain2022}, this module unifies API access across various LLMs and constructs a central \texttt{VerifyAgent}.
Driven by a ReAct reasoning loop~\cite{yao2022react}, \texttt{VerifyAgent} translates workflow task requirements into concrete tool-call sequences to perform operations such as specification analysis, code generation, and test execution.

In addition to common tools such as file-system operations, the middleware integrates custom tools via the Model Context Protocol (MCP)~\cite{mcp2024}.
These specialized tools fall into two primary categories:
\begin{itemize}[topsep=0pt, partopsep=0pt]
    \item \textbf{Workflow control tools.} These tools expose the workflow management capabilities to the LLM, allowing it to query the current stage, trigger checks, and transition between stages.
    \item \textbf{Verification environment tools.} These tools encapsulate underlying operations, enabling the LLM to compile designs, run tests, and retrieve simulation results.
\end{itemize}

By adhering to the MCP standard, these tools can also be exposed as service interfaces.
This interoperability allows external coding agents (e.g., Claude Code, Codex) to access UCAgent's infrastructure, facilitating capability reuse and cross-agent collaboration.

Operationally, the middleware executes tasks through an iterative loop: \textit{context fetching $\rightarrow$ LLM reasoning $\rightarrow$ tool execution $\rightarrow$ environment feedback}.
In each iteration, the LLM analyzes current stage requirements and historical results to determine the next action.
If an error or failure is reported, the middleware captures this feedback and triggers a retry, allowing the LLM to dynamically adjust its strategy rather than strictly following a static script.
Furthermore, the middleware supports human-in-the-loop intervention at critical steps, balancing automation efficiency with human controllability.

\subsection{Workflow Management Module}
\noindent The \textit{Workflow Management Module} orchestrates the execution sequence of the configurable 31-stage workflow.
It strictly gates the progression of the LLM: the workflow can only advance to the subsequent stage when the current stage's outputs pass all constraints enforced by the \textit{Stage Validation Module}.

To ensure flexibility, the workflow is defined via a user-configurable YAML format.
Each stage is independently configured with specific attributes, such as task descriptions, explicit output requirements, and bound checkers, natively supporting nested sub-stages for complex verification tasks.

A critical design principle of this module is distinguishing stage-level goals from tool-level execution.
While the agent-level ReAct loop manages the retries of individual tool calls, the workflow checkers validate whether the generated artifacts strictly satisfy the stage's objectives before allowing the workflow to advance.

\begin{figure}[h]
    \centering
    \includegraphics[width=\textwidth]{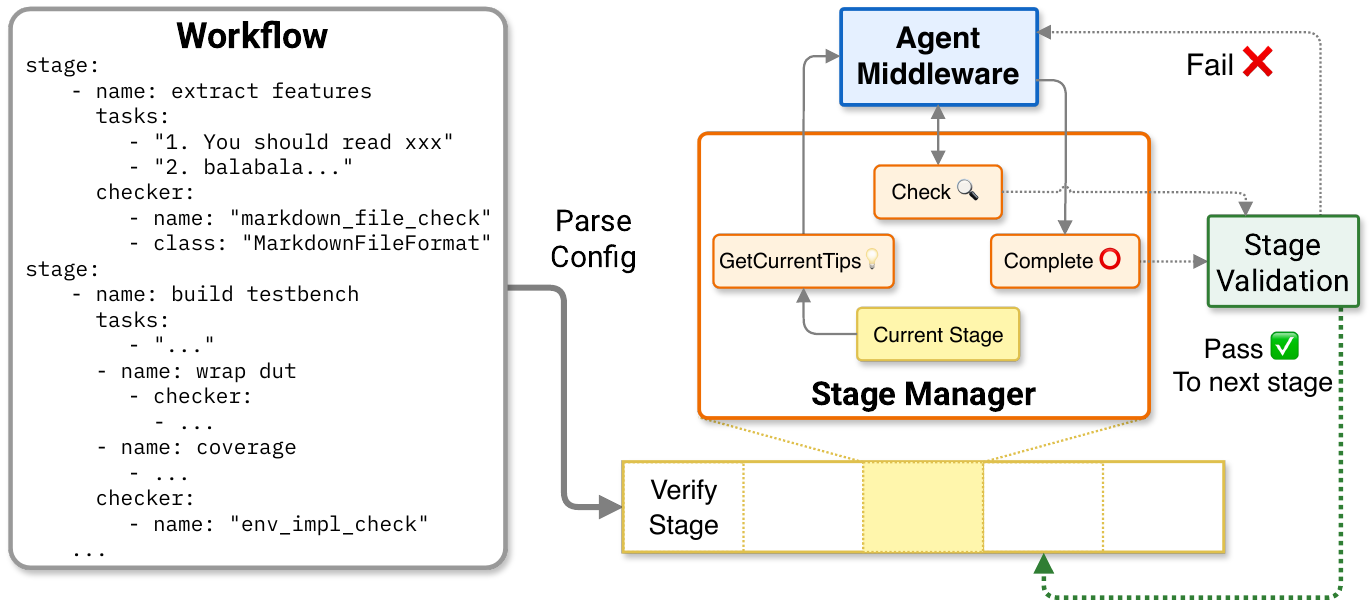}
    \caption{Workflow management module}
    \label{fig:stage-manager}
\end{figure}

As illustrated in Fig.~\ref{fig:stage-manager}, the module has two core classes: \textit{VerifyStage} and \textit{StageManager}.

\begin{itemize}[topsep=0pt, partopsep=0pt, itemsep=0pt, parsep=0.5em]
    \item \textbf{VerifyStage} encapsulates the definition of a single verification node, integrating fields such as the stage name, task description, target output files, and corresponding checker instances.
    \item \textbf{StageManager} acts as the centralized workflow engine. Upon initialization, it parses the YAML configuration into a sequence of \texttt{VerifyStage} objects. During execution, it tracks the active stage index and processes stage-transition requests triggered by the LLM.
\end{itemize}

To interface seamlessly with the \textit{Agent Middleware}, the workflow manager exposes its control capabilities as standard tool calls.
Tools such as \texttt{GetCurrentTips} (retrieve current task context), \texttt{Check} (trigger stage validation), and \texttt{Complete} (verify completion and advance) allow the LLM to actively perceive and navigate its progress without exposing the underlying scheduling implementation.

Furthermore, the module implements the state persistence.
At every stage transition, the \texttt{StageManager} serializes the current progress and execution telemetry (e.g., failure counts and execution time) into a state file.
This mechanism not only permits interrupted tasks to resume precisely from the last completed stage but also provides the empirical data required for workflow refinement.
As previously discussed, stages exhibiting high failure rates in this telemetry indicate that the LLM's task boundaries are insufficiently constrained, guiding engineers to further subdivide or reorganize the workflow architecture.

\subsection{Stage Validation Module}
\noindent The \textit{Stage Validation Module} consists of a collection of extensible checkers.
Each checker automatically evaluates the LLM's output for a given verification stage to ensure the deliverables meet predefined requirements.

UCAgent provides a base checker class, enabling users to implement custom checkers tailored to specific needs. This supports various validation paradigms, including code-based automatic checks, LLM-based peer reviews, and human-in-the-loop judgments.
As shown in Fig.~\ref{fig:overview}, the default workflow integrates checkers covering document formatting, testbench compilation, coverage models, and test reports.
Furthermore, the module includes manual intervention checkers that pause the workflow, allowing verification engineers to review the results of critical stages before proceeding.

\subsubsection{Implementation of VCLM}
\noindent To enforce cross-stage consistency, the Verification Consistency Labeling Mechanism (VCLM) employs a three-level hierarchical label structure:

\begin{itemize}[topsep=0pt, partopsep=0pt]
    \item \textbf{Function Group (\texttt{<FG->}).} It represents top-level categories of verification scenarios in the design specification. For instance, a floating-point ALU might define \texttt{<FG-ARITHMETIC>} for standard operations and \texttt{<FG-WIDEN-ARITHMETIC>} for widening operations.
    \item \textbf{Function Checkpoint (\texttt{<FC->}).} It denotes concrete functional features within a group. Under \texttt{<FG-ARITHMETIC>}, examples include \texttt{<FC-VFADD>} (vector FP addition) and \texttt{<FC-VFSUB>} (vector FP subtraction).
    \item \textbf{Check Point (\texttt{<CK-*>}).} It defines specific verification conditions that must be covered. Under \texttt{<FC-VFADD>}, one might specify \texttt{<CK-FP32>} (FP32-precision addition), \texttt{<CK-FP16>} (FP16-precision addition), and \texttt{<CK-BF16>} (BF16-precision addition).
\end{itemize}

Because VCLM spans multiple workflow stages, it is implemented across a series of stage-specific checkers. Each checker parses the labels generated in its current stage and cross-references them against the label hierarchy established in preceding stages.

\begin{figure}[h]
    \centering
    \includegraphics[width=\linewidth]{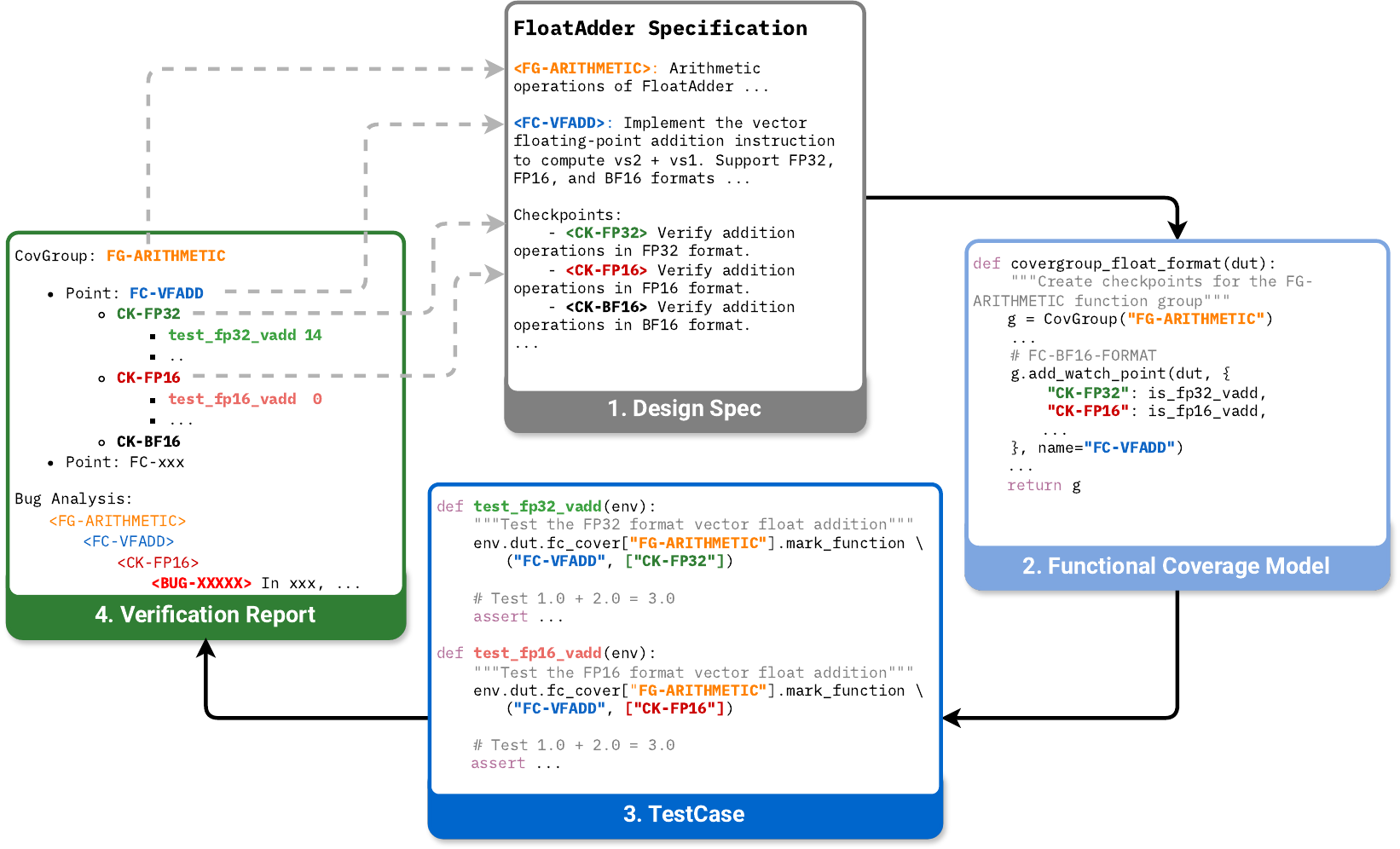}
    \caption{Results of verifying a vector floating-point addition module with VCLM}
    \label{fig:vcmm-tags}
\end{figure}

As illustrated in Fig.~\ref{fig:vcmm-tags}, a typical VCLM-enabled workflow operates as follows:
\begin{itemize}[topsep=0pt, partopsep=0pt]
    \item \textbf{Specification Analysis Stage.} The LLM assigns unique labels to each identified function point within the specification document. For example, a description of FP32 vector addition must be explicitly annotated with \texttt{<FG-ARITHMETIC>}, \texttt{<FC-VFADD>}, and \texttt{<CK-FP32>}. The checker validates label syntax (naming conventions and hierarchical completeness) and ensures a reasonable quantity (e.g., at least one function group).
    \item \textbf{Coverage Modeling Stage.} When generating functional coverage definitions, the LLM must reuse the exact labels from the specification to name each \texttt{CoverGroup}, \texttt{CoverPoint}, and \texttt{Bin}. The checker parses the covergroup hierarchy, extracts the labels, and performs a bidirectional comparison against the specification's label list. It flags labels present in the code but undefined in the specification (indicating hallucinations or naming mistakes), as well as labels defined in the specification but missing in the code (indicating implementation omissions).
    \item \textbf{Test Implementation Stage.} During testcase generation, the LLM utilizes Toffee's \texttt{mark\_function} to explicitly declare which check points the current test function verifies. 
    Post-simulation, the checker parses the execution report to validate that: (1) all marked check points exist in the specification document; (2) all check points defined in the specification are exercised by the tests; and (3) any check points associated with failing testcases are correctly logged in the bug analysis document.
\end{itemize}

\subsection{Python-Based Verification Execution Environment}
\noindent The \textit{Verification Execution Environment} provides the runtime infrastructure for simulating the hardware design entirely within Python. Building upon the \textit{PyDUT} boundary established in the infrastructure layer, this environment hosts the native Python testbench constructed dynamically by the LLM.

To systematically construct and execute this testbench, UCAgent guides the LLM through a highly constrained generation pipeline. 

Initially, the LLM leverages the \texttt{toffee} framework to encapsulate \textit{PyDUT} into high-level procedural functions and generates unit tests to validate these wrappers.
This step effectively acts as an automated smoke test; if these foundational tests persistently fail despite LLM self-correction, the environment flags the associated DUT functionality as a potential hardware bug.
Following this, the LLM generates a functional coverage model using Toffee, mapping it directly to the predefined specification hierarchy.

Finally, under the combined constraints of execution checkers and the VCLM, the LLM generates complex verification scenarios to drive the wrapped DUT.
Each testcase explicitly links executed stimuli to target coverage points via Toffee, ensuring complete traceability.

During the execution phase, the environment utilizes Pytest to run the testcases.
It automatically extracts multi-dimensional execution metrics, including test pass rates, code coverage, and functional coverage.
These metrics are fed back to the LLM as explicit environmental observations.
This data-driven feedback loop enables the LLM to iteratively augment testcases to achieve coverage closure.
Finally, the environment aggregates all execution metrics and verification artifacts to generate a comprehensive verification report.

\section{Evaluation}

\subsection{Experimental Setup}
To evaluate the efficacy of UCAgent, we select five hardware modules, detailed in Table~\ref{tab:subjects}. 
Among these, \texttt{ALU754}, \texttt{IntegerDivider}, and \texttt{ICache-Waylookup} serve as pre-verified baselines, whereas \texttt{LaneFAdd} and \texttt{PageTableWalker} are previously unverified designs.

All subjects are evaluated using UCAgent's default automated workflow. For each module, the LLM is provisioned with the following inputs:
\begin{itemize}[topsep=0pt, partopsep=0pt]
    \item UCAgent's built-in prompt templates.
    \item Design documents for each module.
    \item Verilog files for each module. Except for ALU754 and UART-16550, other modules are implemented in Scala, so corresponding Scala files are also provided.
\end{itemize}

Excluding the PageTableWalker, we evaluate with the following models:
\begin{itemize}[topsep=0pt, partopsep=0pt]
    \item \texttt{claude-sonnet-4-5-20250929}
    \item \texttt{gpt-5-2025-08-07}
    \item \texttt{qwen3-coder-plus-2025-09-23}
\end{itemize}
For all models, \texttt{TopP} and temperature use default settings.\footnote{See: \href{https://platform.claude.com/docs/en/api/ruby/completions/create}{Claude Docs}, \href{https://platform.openai.com/docs/api-reference/responses/create}{OpenAI Docs}, and \href{https://help.aliyun.com/zh/model-studio/qwen-api-reference}{Qwen API Reference}.}

\newcolumntype{Y}{>{\centering\arraybackslash}X}

\begin{table}[h]
    \centering
    \caption{Modules verified with UCAgent}
    \begin{tabularx}{\linewidth}{c c Y}
        \toprule
        Module & LOC & Description \\
        \midrule
        UART 16550 & 268 & Universal Asynchronous Receiver/Transmitter \\
        ALU754 & 572 & IEEE-754 single-precision floating-point unit implemented with combinational logic \\
        \addlinespace
        IntegerDivider & 1024 & RV32M integer divider based on radix-4 SRT algorithm \\
        \addlinespace
        LaneFAdd & 1283 & RVV-compliant vector floating-point unit supporting FP32, FP16, and BF16 \\
        \addlinespace
        ICache-Waylookup & 3483 & Caches metadata queried from MetaArray and ITLB by IPrefetchPipe in the ICache for use by MainPipe \\
        \addlinespace
        PageTableWalker & 155408 & Handles L1/L2 page table access, performs step-by-step page table walks for virtual-to-physical translation, and forwards L3 page table access requests downstream (Last Page Table Walker) \\
        \bottomrule
    \end{tabularx}
    \label{tab:subjects}
\end{table}

\subsection{Overall Verification Effectiveness}

\begin{figure}[h]
    \centering
    \includegraphics[width=1\linewidth]{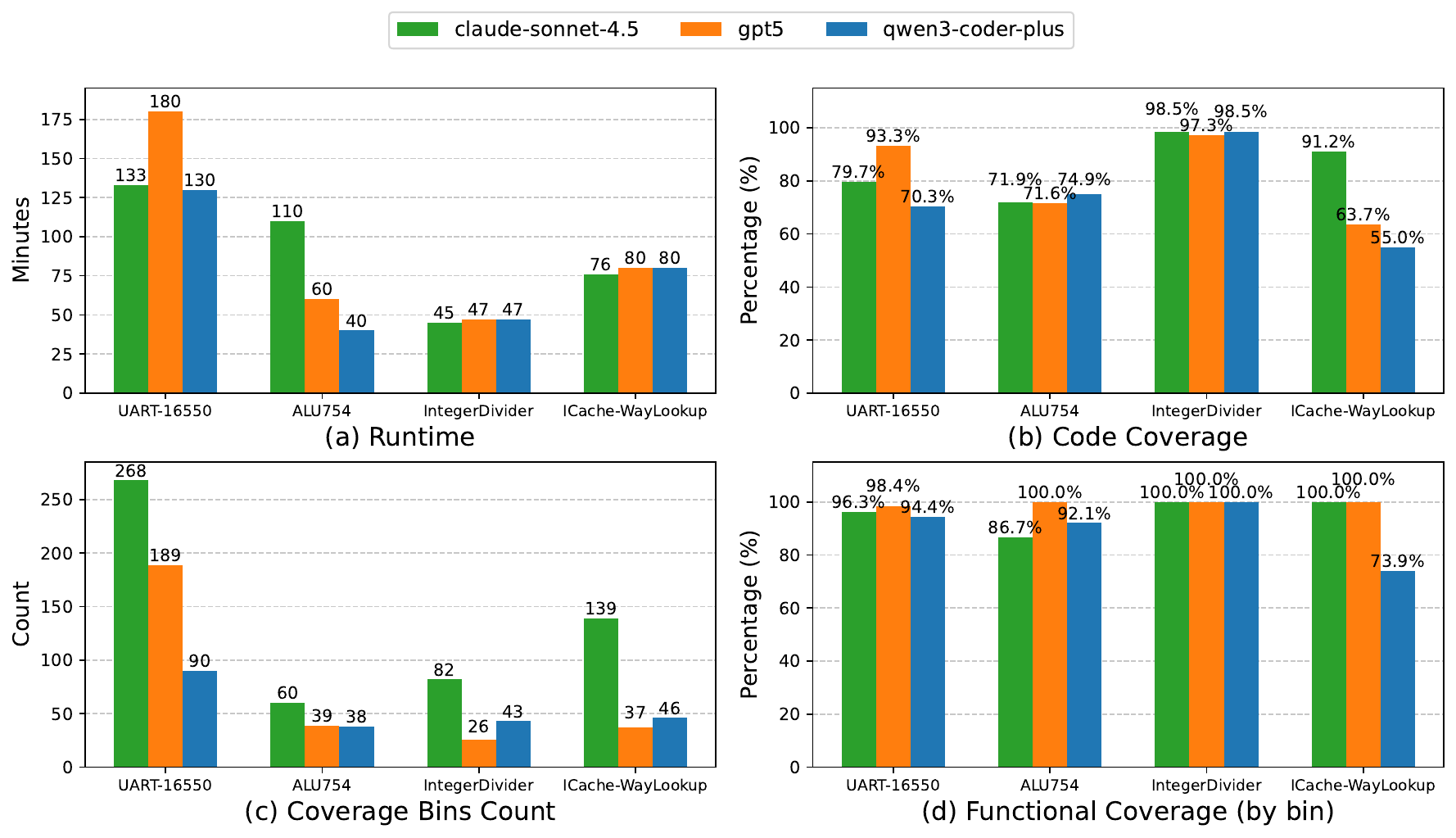}
    \caption{UCAgent evaluation results}
    \label{fig:run-results}
\end{figure}

UCAgent successfully completes end-to-end automated verification for four modules.
As shown in Fig.~\ref{fig:run-results}, execution times vary significantly across models, ranging from 40 to 180 minutes. Notably, despite their smaller codebases, UART-16550 and ALU754 entail complex temporal and combinatorial interaction logic, thus requiring longer verification times.
Conversely, the workflows for IntegerDivider and ICache-WayLookup are highly efficient, converging within 80 minutes for most models.

Regarding code coverage, UCAgent performs optimally on IntegerDivider, exceeding 97\% across all three models. Coverage for UART-16550 and ALU754 typically stabilizes in the 70\%--80\% range.
For the highly complex ICache-WayLookup, coverage outcomes exhibit greater variance (55\%--91\%).

In evaluating functional coverage, UCAgent demonstrates robust overall performance.
IntegerDivider consistently achieves 100\% functional coverage, while ALU754 and UART-16550 generally surpass 86\%.
In particular, Claude-Sonnet-4.5 generates 268 distinct coverage bins for UART-16550, significantly outperforming the other models and showcasing its superior capability for fine-grained functional decomposition.
Overall, the three models exhibit distinct capability profiles:

\begin{itemize}[topsep=0pt, partopsep=0pt]
    \item \textbf{Claude-Sonnet-4.5} achieves the highest average code coverage (85.9\%) and finest functional granularity (137 bins on average).
    However, it exhibits lower initial efficiency (averaging 91 minutes) and encounters the highest failure frequency during testcase generation.
    \item \textbf{GPT-5} demonstrates the most stable functional coverage (consistently 100\%) with minimal variance in stage-wise failures.
    While its overall performance is balanced, it occasionally fluctuates during coverage modeling and adopts a more conservative functional decomposition granularity (73 bins on average).
    \item \textbf{Qwen3-Coder-Plus} excels in execution efficiency (averaging 74 minutes) and exhibits remarkable stability during component and coverage-model generation.
    Nonetheless, this speed trades off against thoroughness, yielding the lowest average code coverage (74.7\%) and functional coverage (90.1\%).
\end{itemize}

\subsection{Stage-wise Execution Analysis}

\begin{figure}[h]
    \centering
    \includegraphics[width=1\linewidth]{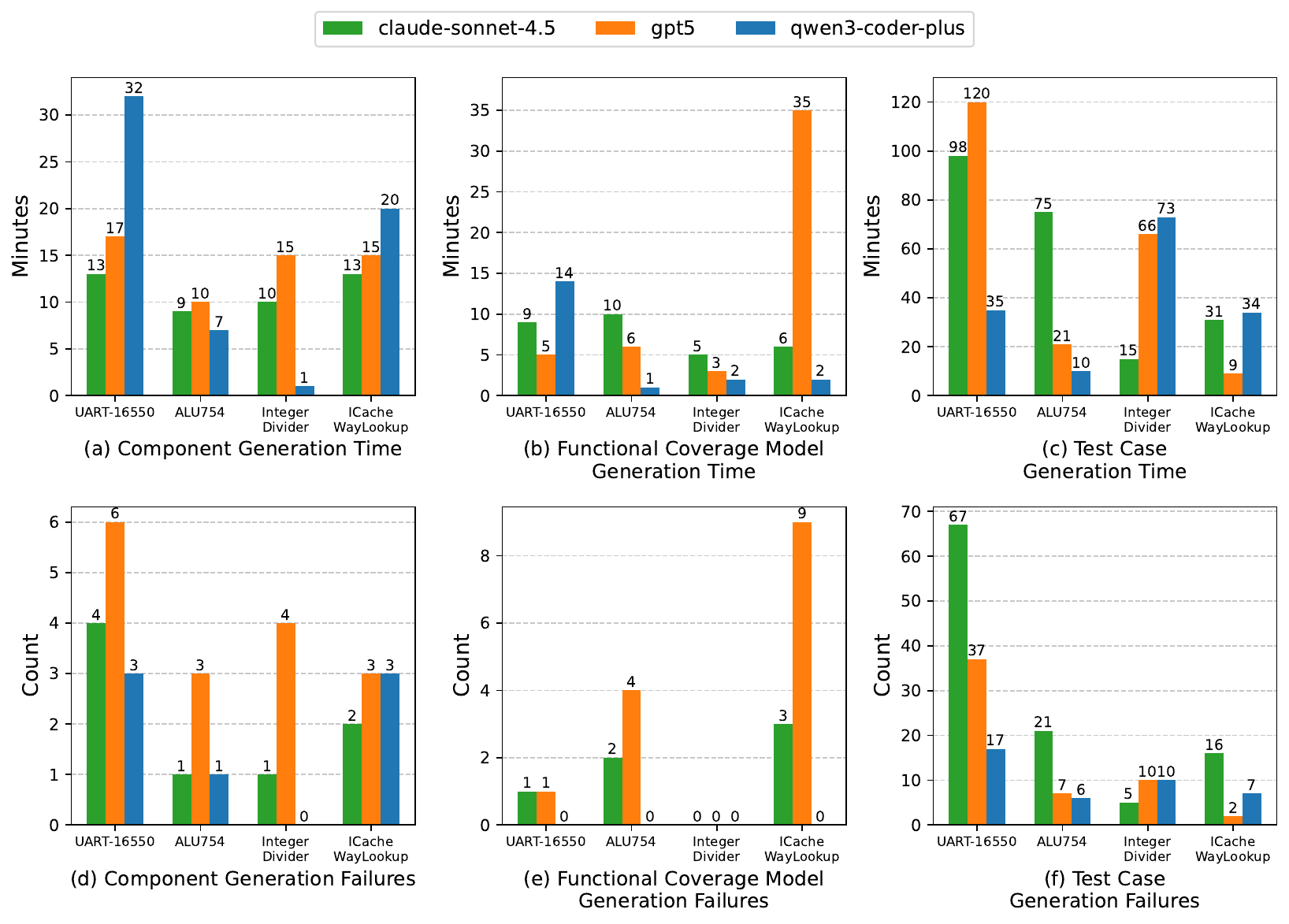}
    \caption{Time and failure counts of each stage when UCAgent builds the verification environment}
    \label{fig:detail-run-results}
\end{figure}

Fig.~\ref{fig:detail-run-results} dissects the verification workflow into three primary phases: component generation, coverage-model generation, and testcase generation.
The temporal distribution and failure rates across these stages vividly illuminate the inherent complexities of distinct verification tasks.

\textbf{Component Generation} exhibits stable execution times, typically concluding within 10--20 minutes. 
Failure counts remain consistently low (1--6), indicating that UCAgent's underlying checker mechanism effectively constrains the LLMs to produce structurally compliant verification components.

\textbf{Coverage-Model Generation} generally finishes under 10 minutes as the shortest phase.
Qwen3-Coder-Plus proves exceptionally efficient here, requiring merely 1--2 minutes for IntegerDivider and ICache-WayLookup.
The near-zero failure rates strongly suggest that the VCLM explicitly provides clear, structured semantic constraints, effectively mitigating LLM hallucinations.

\textbf{Testcase Generation} incurs the highest failure rates as the most resource-intensive phase.
This is expected, as valid testcases must simultaneously satisfy strict syntactic correctness, functional assertions, and targeted coverage annotations.
For instance, Claude-Sonnet-4.5 encounters 67 iterative failures on UART-16550 yet ultimately achieves verification closure, powerfully demonstrating the robustness of UCAgent's checker-driven self-correction loop.

By combining the stage-level data in Fig.~\ref{fig:detail-run-results}, we can further explain the model capability profiles above:
\begin{itemize}[topsep=0pt, partopsep=0pt]
    \item \textbf{Claude-Sonnet-4.5} relies heavily on iterative refinement during test generation (averaging 27.3 failures), explaining its longer runtime. However, this sustained trial-and-error approach directly yields the highest code coverage and most comprehensive test suites.
    \item \textbf{GPT-5} maintains stage-wise equilibrium but displays occasional volatility during coverage modeling (0--9 failures), though its self-correction ensures this does not degrade final functional coverage.
    \item \textbf{Qwen3-Coder-Plus}'s efficiency stems from a coarser functional decomposition strategy. Generating fewer coverage bins inherently reduces the burden on subsequent testcase generation, resulting in fewer failures and faster execution, albeit at the cost of lower final coverage.
\end{itemize}

\subsection{Case Studies}
To validate UCAgent's real-world applicability, we conducted experiments on specific modules, highlighting its capability on both established designs (IntegerDivider) and unverified hardware (LaneFAdd, PageTableWalker).

\subsubsection{IntegerDivider Verification (Domain Knowledge Boundaries)}
Our evaluation reveals significant disparities in the inherent domain knowledge embedded within different LLMs. 
During testcase generation for the RV32M integer divider, Qwen3-Coder-Plus repeatedly failed to generate the correct expected behavior for the specific boundary case where \textbf{the dividend is the minimum negative value ($-2^{31}$) and the divisor is $-1$}, despite this overflow behavior being explicitly defined in the RISC-V specification.
Conversely, Claude-Sonnet-4.5 and GPT-5 robustly identified and covered this edge case.
This highlights that while UCAgent orchestrates the workflow, absolute verification completeness remains bounded by the base model's domain knowledge, suggesting future needs for targeted knowledge injection.

In contrast, Claude-Sonnet-4.5 and GPT-5 are more robust and cover the key boundary case above. These results suggest that knowledge boundaries of LLMs may limit verification completeness, and techniques such as prompt enhancement or domain knowledge injection are needed.

\subsubsection{LaneFAdd Verification (Automated Bug Discovery)}
Driven by \texttt{claude-sonnet-4-5-20250929}, UCAgent successfully found three previously undiscovered boundary-condition bugs involving special-value handling under FP16 and BF16 precision within the unverified LaneFAdd module.
Crucially, by leveraging the VCLM, UCAgent automatically traced these failures back to their specific function and check points, seamlessly facilitating rapid localization and resolution by hardware developers.

\subsubsection{PageTableWalker Verification (Human-in-the-Loop Applicability)}
Due to the massive scale of this module (155,408 lines), UCAgent could not complete the verification task using the default workflow. We therefore adopted a human-in-the-loop verification strategy and achieved the expected verification outcome within 8 hours.

\section{Discussion}

\subsection{Effectiveness}
Experimental results demonstrate that UCAgent can autonomously complete the end-to-end unit-level chip verification workflow.
This capability fundamentally stems from the structural synergy among its four core design mechanisms:
\begin{enumerate}
    \item \textbf{Python-based verification implementation} reduces the difficulty of generating verification code and allows reusing LLM strengths from software engineering;
    \item \textbf{Checker mechanism} validates LLM behavior through deterministic constraint logic and keeps stage failure counts within an acceptable range;
    \item \textbf{Configurable workflow} adapts verification strategies to module characteristics and meets diverse needs across module complexity;
    \item \textbf{VCLM} maintains verification consistency via a labeling system and ensures traceability of results.
\end{enumerate}

\subsection{Limitations}
Despite its demonstrated efficacy, UCAgent exhibits several limitations that inform future research directions:
\begin{itemize}[topsep=0pt, partopsep=0pt]
    \item For massive industrial-scale designs, establishing hierarchical workflow configurations and modular decomposition still necessitates manual engineering effort.
    \item Testcase generation continues to incur the highest failure and retry rates, which can lead to prolonged iteration cycles when confronting highly complex verification scenarios.
    \item The system's verification completeness remains bounded by the base LLM's intrinsic domain knowledge. 
    When models misunderstand specific hardware specifications, UCAgent currently lacks systematic mechanisms (such as dynamic knowledge injection) to compensate for these gaps.
    \item The current 26-stage workflow architecture is statically predefined by human experts.
    Future iterations could explore fully autonomous, LLM-driven dynamic task decomposition to further elevate the level of automation.
\end{itemize}

\printbibliography

@misc {daniel2023state,
    title = "The State of IC and ASIC Functional Verification",
    author = {Daniel Payne},
    year = 2023,
    url = "https://semiwiki.com/eda/324443-the-state-of-ic-and-asic-functional-verification",
    note = "[Online; accessed 2025-09-29]"
}

@article{molina2007functional,
  title={Functional verification: Approaches and challenges},
  author={Molina, A and Cadenas, Oswaldo},
  journal={Latin American applied research},
  volume={37},
  number={1},
  pages={65--69},
  year={2007},
  publisher={SciELO Argentina}
}

@article{foster2022fcstudy,
  title={The 2022 Wilson Research Group Functional Verification Study},
  author={Foster, Harry and others},
  journal={Siemens.com},
  year={2022}
}

@article{foster2024fcstudy,
  title={2024 Wilson Research Group IC/ASIC functional verification trend report},
  author={Foster, Harry and others},
  journal={Siemens.com},
  year={2024}
}

@Inbook{Mehta2018,
author="Mehta, Ashok B.",
title="Constrained Random Verification (CRV)",
bookTitle="ASIC/SoC Functional Design Verification: A Comprehensive Guide to Technologies and Methodologies",
year="2018",
publisher="Springer International Publishing",
address="Cham",
pages="65--74",
isbn="978-3-319-59418-7",
doi="10.1007/978-3-319-59418-7_5",
}

@article{kern1999formal,
  title={Formal verification in hardware design: a survey},
  author={Kern, Christoph and Greenstreet, Mark R},
  journal={ACM Transactions on Design Automation of Electronic Systems (TODAES)},
  volume={4},
  number={2},
  pages={123--193},
  year={1999},
  publisher={ACM New York, NY, USA}
}

@techreport{foster2025bottleneck,
  title     = {Breaking the bottleneck: Overcoming the Verification Productivity Gap 2.0},
  author    = {Foster, Harry D.},
  year      = {2025},
  institution = {Siemens Digital Industries Software},
  type      = {White Paper},
  url       = {https://resources.sw.siemens.com/en-US/white-paper-breaking-the-bottleneck-overcoming-the-verification-productivity-gap-2-0/}
}

@online{synopsys_vso_ai,
  author       = {Synopsys},
  title        = {VSO.ai: AI-Driven Verification Space Optimization},
  year         = {2024},
  url          = {https://www.synopsys.com/ai/ai-powered-eda/vso-ai.html},
  urldate      = {2025-11-26}
}

@online{synopsys_ai,
  author       = {Synopsys},
  title        = {Synopsys.ai: Full-Stack AI-Powered EDA Suite},
  year         = {2024},
  url          = {https://www.synopsys.com/ai/ai-powered-eda.html},
  urldate      = {2025-11-26}
}

@online{cadence_verisium,
  author       = {Cadence Design Systems},
  title        = {Verisium: AI-Driven Verification Platform},
  year         = {2024},
  url          = {https://www.cadence.com/en_US/home/tools/system-design-and-verification/ai-driven-verification.html},
  urldate      = {2025-11-26}
}

@online{cadence_verisium_debug,
  author       = {Cadence Design Systems},
  title        = {Verisium Debug: Machine-Learning Accelerated Debugging},
  year         = {2024},
  url          = {https://www.cadence.com/en_US/home/tools/system-design-and-verification/ai-driven-verification/verisium-debug.html},
  urldate      = {2025-11-26}
}

@online{cadence_jedai,
  author       = {Cadence Design Systems},
  title        = {JedAI: Big-Data and AI Platform for EDA},
  year         = {2024},
  url          = {https://www.cadence.com/en_US/home/tools/ai/jedai-platform.html},
  urldate      = {2025-11-26}
}

@online{siemens_eda_ai_system,
  author       = {Siemens EDA},
  title        = {EDA AI System: Generative and Agentic AI for Chip Design and Verification},
  year         = {2024},
  url          = {https://eda.sw.siemens.com/en-US/ai/eda-ai-system/},
  urldate      = {2025-11-26}
}

@online{siemens_solido_ai,
  author       = {Siemens EDA},
  title        = {Solido: Generative and Agentic AI for Custom IC Verification},
  year         = {2024},
  url          = {https://eda.sw.siemens.com/en-US/ic/solido/},
  urldate      = {2025-11-26}
}

@misc{lozhkov2024starcoder,
      title={StarCoder 2 and The Stack v2: The Next Generation}, 
      author={Anton Lozhkov and Raymond Li and Loubna Ben Allal and Federico Cassano and Joel Lamy-Poirier and Nouamane Tazi and Ao Tang and Dmytro Pykhtar and Jiawei Liu and Yuxiang Wei and Tianyang Liu and Max Tian and Denis Kocetkov and Arthur Zucker and Younes Belkada and Zijian Wang and Qian Liu and Dmitry Abulkhanov and Indraneil Paul and Zhuang Li and Wen-Ding Li and Megan Risdal and Jia Li and Jian Zhu and Terry Yue Zhuo and Evgenii Zheltonozhskii and Nii Osae Osae Dade and Wenhao Yu and Lucas Krauß and Naman Jain and Yixuan Su and Xuanli He and Manan Dey and Edoardo Abati and Yekun Chai and Niklas Muennighoff and Xiangru Tang and Muhtasham Oblokulov and Christopher Akiki and Marc Marone and Chenghao Mou and Mayank Mishra and Alex Gu and Binyuan Hui and Tri Dao and Armel Zebaze and Olivier Dehaene and Nicolas Patry and Canwen Xu and Julian McAuley and Han Hu and Torsten Scholak and Sebastien Paquet and Jennifer Robinson and Carolyn Jane Anderson and Nicolas Chapados and Mostofa Patwary and Nima Tajbakhsh and Yacine Jernite and Carlos Muñoz Ferrandis and Lingming Zhang and Sean Hughes and Thomas Wolf and Arjun Guha and Leandro von Werra and Harm de Vries},
      year={2024},
      eprint={2402.19173},
      archivePrefix={arXiv},
      primaryClass={cs.SE}
}

@inproceedings{tsai2024rtlfixer,
  title={Rtlfixer: Automatically fixing rtl syntax errors with large language model},
  author={Tsai, YunDa and Liu, Mingjie and Ren, Haoxing},
  booktitle={Proceedings of the 61st ACM/IEEE Design Automation Conference},
  pages={1--6},
  year={2024}
}

@article{zhang2025understanding,
  title={Understanding and Mitigating Errors of LLM-Generated RTL Code},
  author={Zhang, Jiazheng and Liu, Cheng and Li, Huawei},
  journal={arXiv preprint arXiv:2508.05266},
  year={2025}
}

@inproceedings{cui2024origen,
  title={Origen: Enhancing rtl code generation with code-to-code augmentation and self-reflection},
  author={Cui, Fan and Yin, Chenyang and Zhou, Kexing and Xiao, Youwei and Sun, Guangyu and Xu, Qiang and Guo, Qipeng and Liang, Yun and Zhang, Xingcheng and Song, Demin and others},
  booktitle={Proceedings of the 43rd IEEE/ACM International Conference on Computer-Aided Design},
  pages={1--9},
  year={2024}
}

@article{zhao2025codev,
  title={Codev: Empowering llms with hdl generation through multi-level summarization},
  author={Zhao, Yang and Huang, Di and Li, Chongxiao and Jin, Pengwei and Song, Muxin and Xu, Yinan and Nan, Ziyuan and Gao, Mingju and Ma, Tianyun and Qi, Lei and others},
  journal={IEEE Transactions on Computer-Aided Design of Integrated Circuits and Systems},
  year={2025},
  publisher={IEEE}
}

@inproceedings{ho2025verilogcoder,
  title={Verilogcoder: Autonomous verilog coding agents with graph-based planning and abstract syntax tree (ast)-based waveform tracing tool},
  author={Ho, Chia-Tung and Ren, Haoxing and Khailany, Brucek},
  booktitle={Proceedings of the AAAI Conference on Artificial Intelligence},
  volume={39},
  number={1},
  pages={300--307},
  year={2025}
}

@article{zhu2025codev,
  title={CodeV-R1: Reasoning-Enhanced Verilog Generation},
  author={Zhu, Yaoyu and Huang, Di and Lyu, Hanqi and Zhang, Xiaoyun and Li, Chongxiao and Shi, Wenxuan and Wu, Yutong and Mu, Jianan and Wang, Jinghua and Zhao, Yang and others},
  journal={arXiv preprint arXiv:2505.24183},
  year={2025}
}

@article{pinckney2025comprehensive,
  title={Comprehensive Verilog Design Problems: A Next-Generation Benchmark Dataset for Evaluating Large Language Models and Agents on RTL Design and Verification},
  author={Pinckney, Nathaniel and Deng, Chenhui and Ho, Chia-Tung and Tsai, Yun-Da and Liu, Mingjie and Zhou, Wenfei and Khailany, Brucek and Ren, Haoxing},
  journal={arXiv preprint arXiv:2506.14074},
  year={2025}
}

@article{ye2025uvm2,
  title={From Concept to Practice: an Automated LLM-aided UVM Machine for RTL Verification},
  author={Ye, Junhao and Hu, Yuchen and Xu, Ke and Pan, Dingrong and Chen, Qichun and Zhou, Jie and Zhao, Shuai and Fang, Xinwei and Wang, Xi and Guan, Nan and others},
  journal={arXiv preprint arXiv:2504.19959v3},
  year={2025}
}

@article{liu2025multi,
  title={A Multi-Agent Generative AI Framework for IC Module-Level Verification Automation},
  author={Liu, Wenbo and Hou, Forbes and Zhang, Jon and Liu, Hong and Lei, Allen},
  journal={arXiv preprint arXiv:2507.21694},
  year={2025}
}

@article{zhao2025pro,
  title={PRO-V: An Efficient Program Generation Multi-Agent System for Automatic RTL Verification},
  author={Zhao, Yujie and Wu, Zhijing and Zhang, Hejia and Yu, Zhongming and Ni, Wentao and Ho, Chia-Tung and Ren, Haoxing and Zhao, Jishen},
  journal={arXiv preprint arXiv:2506.12200},
  year={2025}
}

@inproceedings{sun2023nl2sva,
  title={Towards Improving Verification Productivity with Circuit-Aware Translation of Natural Language to SystemVerilog Assertions},
  author={Chuyue Sun and Christopher Hahn and Caroline Trippel},
  booktitle={International Workshop on Deep Learning-aided Verification (DAV)},
  year={2023}
}

@inproceedings{liu2024domainadapted,
  title={Domain-Adapted LLMs for VLSI Design and Verification: A Case Study on Formal Verification},
  author={Mingjie Liu and Minwoo Kang and Ghaith Bany Hamad and Syed Suhaib and Haoxing Ren},
  booktitle={VLSI Test Symposium (VTS)},
  year={2024}
}

@article{rahul2024llmassertions,
  title={LLM-Assisted Generation of Hardware Assertions},
  author={Rahul Kande and Hammond Pearce and Benjamin Tan and Brendan Dolan-Gavitt and Shailja Thakur and Ramesh Karri and Jeyavijayan Rajendran},
  journal={arXiv preprint arXiv:2306.14027},
  year={2024}
}

@article{pulavarthi2025ready,
  title={Are LLMs Ready for Practical Adoption for Assertion Generation?},
  author={Vaishnavi Pulavarthi and Deeksha Nandal and Soham Dan and Debjit Pal},
  journal={arXiv preprint arXiv:2502.20633},
  year={2025}
}

@article{bhandari2024fsmtestbench,
  title={LLM-Aided Testbench Generation and Bug Detection for Finite-State Machines},
  author={Jitendra Bhandari and Johann Knechtel and Ramesh Narayanaswamy and Siddharth Garg and Ramesh Karri},
  journal={arXiv preprint arXiv:2406.17132},
  year={2024}
}

@article{huang2024verilogassistant,
  title={Towards LLM-Powered Verilog RTL Assistant: Self-Verification and Self-Correction},
  author={Hanxian Huang and Zhenghan Lin and Zixuan Wang and Xin Chen and Ke Ding and Jishen Zhao},
  journal={arXiv preprint arXiv:2406.00115},
  year={2024}
}

@inproceedings{xiao2024neuroverify,
  title={LLM-Based Processor Verification: A Case Study for Neuromorphic Processor},
  author={Chao Xiao and Yifei Deng and Zhijie Yang and Renzhi Chen and Hong Wang and Jingyue Zhao and Huadong Dai and Lei Wang and Yuhua Tang and Weixia Xu},
  booktitle={Design, Automation and Test in Europe Conference (DATE)},
  year={2024}
}

@techreport{hong2025context,
  title = {Context Rot: How Increasing Input Tokens Impacts LLM Performance},
  author = {Hong, Kelly and Troynikov, Anton and Huber, Jeff},
  year = {2025},
  month = {July},
  institution = {Chroma},
  url = {https://research.trychroma.com/context-rot},
}

@misc{enwiki:1316197198,
    author = "{Wikipedia contributors}",
    title = "Hallucination (artificial intelligence) --- {Wikipedia}{,} The Free Encyclopedia",
    year = "2025",
    url = "https://en.wikipedia.org/w/index.php?title=Hallucination_(artificial_intelligence)&oldid=1316197198",
    note = "[Online; accessed 11-October-2025]"
}

@book{osherove2024art,
  title={The Art of Unit Testing: With Examples in JavaScript},
  author={Osherove, Roy and Khorikov, Vladimir and others},
  year={2024},
  publisher={Simon and Schuster}
}

@article{heo2024llms,
    title={Do LLMs" know" internally when they follow instructions?},
    author={Heo, Juyeon and Heinze-Deml, Christina and Elachqar, Oussama and Chan, Kwan Ho Ryan and Ren, Shirley and Nallasamy, Udhay and Miller, Andy and Narain, Jaya},
    journal={arXiv preprint arXiv:2410.14516},
    year={2024}
}

@misc{abdelaziz2024granitefunctioncallingmodelintroducing,
      title={Granite-Function Calling Model: Introducing Function Calling Abilities via Multi-task Learning of Granular Tasks}, 
      author={Ibrahim Abdelaziz and Kinjal Basu and Mayank Agarwal and Sadhana Kumaravel and Matthew Stallone and Rameswar Panda and Yara Rizk and GP Bhargav and Maxwell Crouse and Chulaka Gunasekara and Shajith Ikbal and Sachin Joshi and Hima Karanam and Vineet Kumar and Asim Munawar and Sumit Neelam and Dinesh Raghu and Udit Sharma and Adriana Meza Soria and Dheeraj Sreedhar and Praveen Venkateswaran and Merve Unuvar and David Cox and Salim Roukos and Luis Lastras and Pavan Kapanipathi},
      year={2024},
      eprint={2407.00121},
      archivePrefix={arXiv},
      primaryClass={cs.LG},
      url={https://arxiv.org/abs/2407.00121}, 
}

@misc{langchain2022,
  author       = {Chase, Harrison and {LangChain, Inc.}},
  title        = {{LangChain}: A framework for building applications powered by large language models},
  year         = {2022},
  howpublished = {\url{https://github.com/langchain-ai/langchain}},
  note         = {Available at: https://www.langchain.com/. Accessed: 2026-01-01}
}

@inproceedings{yao2022react,
    title={React: Synergizing reasoning and acting in language models},
    author={Yao, Shunyu and Zhao, Jeffrey and Yu, Dian and Du, Nan and Shafran, Izhak and Narasimhan, Karthik R and Cao, Yuan},
    booktitle={The eleventh international conference on learning representations},
    year={2022}
}

@online{mcp2024,
    title = {Model Context Protocol: Getting Started},
    author = {{Model Context Protocol Authors}},
    year = {2024},
    url = {https://modelcontextprotocol.io/docs/getting-started/intro},
    urldate = {2024-12-19},
    note = {Accessed: 2025-11-11}
}

@article{yang2025multiswebench,
    title={Multi-swe-bench: A multilingual benchmark for issue resolving},
    author={Yang, Shengnan and others},
    journal={arXiv preprint arXiv:2504.02605},
    year={2025}
}

@misc{cocotb_issue3110,
    author = {gruvw},
    title = {Signals not updated after rising edge · Issue \#3110 · cocotb/cocotb},
    year = {2022},
    url = {https://github.com/cocotb/cocotb/issues/3110},
    note = {Accessed: 2025 Apr 09},
}

@article{brown2020language,
  title={Language models are few-shot learners},
  author={Brown, Tom and Mann, Benjamin and Ryder, Nick and Subbiah, Melanie and Kaplan, Jared D and Dhariwal, Prafulla and Neelakantan, Arvind and Shyam, Pranav and Sastry, Girish and Askell, Amanda and others},
  journal={Advances in neural information processing systems},
  volume={33},
  pages={1877--1901},
  year={2020}
}

@manual{intel_xhci_reqspec_2019_rev122,
  title        = {{eXtensible Host Controller Interface for Universal Serial Bus (xHCI) Requirements Specification}},
  organization = {{Intel Corporation}},
  year         = {2019},
  month        = may,
  note         = {Revision 1.22},
  url          = {https://www.intel.com/content/dam/www/public/us/en/documents/technical-specifications/extensible-host-controler-interface-usb-xhci.pdf},
  urldate      = {2026-01-06}
}

@misc{picker,
  author       = {{XS-MLVP}},
  title        = {Picker: Pick your favorite language to verify your chip},
  year         = {2024},
  publisher    = {GitHub},
  howpublished = {\url{https://github.com/XS-MLVP/picker}},
  note         = {Accessed: 2025-01-11}
}

@misc{toffee,
  author       = {{XS-MLVP}},
  title        = {Toffee: A framework for building hardware verification platform using software method},
  year         = {2024},
  publisher    = {GitHub},
  howpublished = {\url{https://github.com/XS-MLVP/toffee}},
  note         = {Accessed: 2025-01-11}
}

\end{document}